\documentclass[twocolumn,letterpaper,aps,prl,superscriptaddress,showpacs,amsmath]{revtex4}

\usepackage{graphicx}

\begin{document}

\title{Orbital order and possible superconductivity in 
LaNiO${}_\mathbf{3}$/LaMO${}_\mathbf{3}$ superlattices}

\author{Ji\v{r}\'{\i} Chaloupka}
\affiliation{Max-Planck-Institut f\"ur Festk\"orperforschung,
Heisenbergstrasse 1, D-70569 Stuttgart, Germany}
\affiliation{Institute of Condensed Matter Physics,
Masaryk University, Kotl\'a\v{r}sk\'a 2, 61137 Brno, Czech Republic}

\author{Giniyat Khaliullin}
\affiliation{Max-Planck-Institut f\"ur Festk\"orperforschung,
Heisenbergstrasse 1, D-70569 Stuttgart, Germany}

\begin{abstract}
A hypothetical layered oxide La$_2$NiMO$_{6}$ where NiO$_2$ and MO$_2$ planes
alternate along the \mbox{$c$-axis} of ABO$_3$ perovskite lattice is
considered theoretically. Here, M denotes a trivalent cation Al,
Ga,\ldots~such that MO$_2$ planes are insulating and suppress the $c$-axis
charge transfer.  We predict that correlated $e_g$ electrons in the NiO$_2$
planes develop a planar $x^2-y^2$ orbital order driven by the reduced
dimensionality and further supported by epitaxial strain from the substrate.
Low energy electronic states can be mapped to a single-band $t-t'-J$ model,
suggesting favorable conditions for high-$T_c$ superconductivity. 
\end{abstract}

\date{\today}


% 71 Electronic structure of bulk materials
% 71.27.+a Strongly correlated electron systems; heavy fermions

% 75 Magnetic properties and materials
% 75.10.-b General theory and models of magnetic ordering 
% 75.30.Et Exchange and superexchange interactions

% 74 Superconductivity
% 74.70.-b Superconducting materials (for cuprates, see 74.72.-h)
% 74.78.Bz High-Tc films
% 74.78.Fk Multilayers, superlattices, heterostructures
% 74.78.-w Superconducting films and low-dimensional structures

\maketitle

Despite decades of extensive research, cuprates remain the only compounds to
date hosting the high-temperature superconductivity (SC). 
On empirical grounds, the key electronic and structural elements that support
high $T_c$ values are well known -- no orbital degeneracy, spin one-half,
quasi two-dimensionality (2D), strong antiferromagnetic (AF) correlations. 
While these properties are partially realized in various materials 
({\it e.g.}, layered cobaltates), only cuprates do possess all of them. 

A unique feature of the high-$T_c$ cuprates is the presence of an extended
doping interval $0.05 \lesssim \delta \lesssim 0.20$ where the correlated electron 
maintains its (plane-wave/localized-particle) duality, and both fermionic and
spin statistics may operate in physically relevant energy scales.
Multifaceted behavior of electrons results in an exotic ``normal'' state of the
cuprates with ill-defined quasiparticles, pseudogap {\it etc}, which condenses
into the superconducting state below $T_c$. There are a number of strongly
correlated metallic oxides \cite{Ima98} based on $S=1/2$ 3d-ions as Ti$^{3+}$
and V$^{4+}$ (both with a single $t_{2g}$ electron), Co$^{4+}$ (a $t_{2g}$
hole) and Ni$^{3+}$ (closed $t_{2g}$ shell plus one $e_g$ electron) that
possess a low-spin state in octahedral environment. These compounds show a
great diversity of physical properties \cite{Ima98}; however, the mysterious
strange-metal phase from which anomalous SC may emerge is missing. 

Apart from dimensionality, the orbital degeneracy is ``to blame'' here.
Originating from high symmetry of the MeO$_6$ octahedron  -- a common building
block of both pseudocubic and layered perovskites, -- the orbital degeneracy
enlarges the Hilbert space and  relaxes kinematical constraints on the
electron motion.  Consequently, a fermionic coherency is enhanced and doping
induced  insulator-metal transitions occur without a reference to the
pseudogap phase.  {\it E.g.}, in La$_{1-x}$Sr$_x$TiO$_3$ the formation of a
three-band, correlated Fermi-liquid completes within just a few percent doping
range near $x\sim 0.05$ \cite{Tok93}. 

The orbital degeneracy strongly reduces AF correlations (believed to be
crucial in cuprate physics), as electrons are allowed to have parallel spins
residing on the different orbitals.  This leads to competing Ferro- and AF-interactions  
that result in a rich variety of magnetic states 
in $S=1/2$ oxides RTiO$_3$, Na$_x$CoO$_2$,
Sr$_2$CoO$_4$, RNiO$_3$, NaNiO$_2$, {\it etc}.  In contrast, spin correlations
in single-band cuprates are of AF nature exclusively and hence strong. 

\begin{figure}[h!]
\includegraphics[scale=0.95]{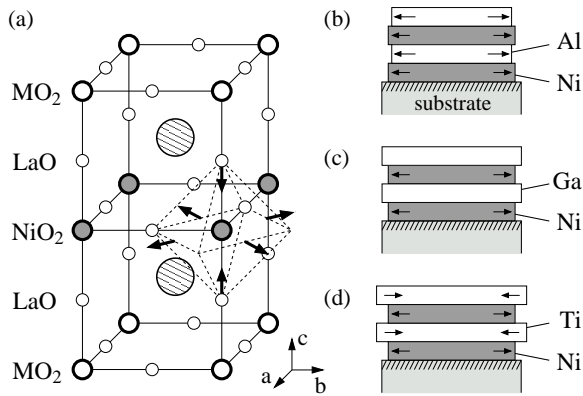}
\caption{
(a) Superlattice La$_2$NiMO$_6$ with alternating NiO$_2$ and MO$_2$ planes.
MO$_2$ layers suppress the $c$-axis hopping resulting in 2D electronic
structure. Arrows indicate the $c$-axis compression of the NiO$_6$ octahedron
imposed by tensile epitaxial strain and supported by Jahn-Teller coupling.
(b,c,d) Strain induced stretching of the NiO$_2$ planes occurs when 
superlattices with M=Al, Ga, Ti respectively are grown on SrTiO$_3$ or
LaGaO$_3$ substrates having large lattice parameter compared to that of
LaNiO$_3$. Expected deformations are indicated by arrows. 
}
\label{fig:structure}
\end{figure}

How to suppress the orbital degeneracy and promote cuprate-like physics in
other $S=1/2$ oxides? In this Letter, we suggest and argue theoretically
that this goal can be achieved in oxide superlattices.  Specifically, we focus
on Ni-based superlattices (see Fig.~\ref{fig:structure}) which can be
fabricated using recent advances in oxide heterostructure technology
(\cite{Izu01,Oht02,Cha06} and references therein).  While the proposed
compound has a pseudocubic ABO$_3$ structure, its low-energy electronic states
are confined to the NiO$_2$ planes hence are of quasi-2D nature.
A substrate induced compression of the NiO$_6$ octahedra further stabilizes the
$x^2-y^2$ orbital. Net effect is a strong enhancement of AF correlations among
spin one-half electrons residing predominantly on a single band of $x^2-y^2$
symmetry. Thus, all the ``high-$T_c$ conditions'' are perfectly met. Moreover,
the presence of a virtual $3z^2-r^2$ orbital deforms the band dispersion
compared to cuprates, leading  to an enhanced next-neighbor hopping $t'$ 
which is known to support higher $T_c$ values~\cite{Pav01}.   

% Fig.1 -------------------------------------------------------------------
%--------------------------------------------------------------------------

The proposed superlattices can be viewed as a layer-by-layer ``mixture'' of a
correlated $e_g$-band metal LaNiO$_3$ and a band insulator LaMO$_3$.  The
MO$_2$ planes with a trivalent M=Al or Ga serve here as block layers
suppressing the $c$-axis hopping.  The lattice parameters of  LaAlO$_3$
($\simeq 3.79\text{\AA}$) and LaGaO$_3$ ($\simeq 3.89\text{\AA}$) are close to
that of LaNiO$_3$ ($\simeq 3.83\text{\AA}$); further,  NiO$_2$ and MO$_2$
planes have the same nominal charges. These factors should result in a minimal
only structural and electronic mismatch, suggesting a stability of
La$_2$NiMO$_6$ compounds.  Yet another intriguing option is the case of M=Ti 
(the lattice parameter of  LaTiO$_3$ $\simeq 3.96\text{\AA}$), 
where TiO$_2$ planes would themselves have a spin
one-half residing on the $t_{2g}$ orbital.  As the $t_{2g}$ shell of Ni$^{3+}$
is full while $e_g$ level of Ti$^{3+}$ is located well above the Fermi energy
(roughly at $10Dq \sim 2\:\mathrm{eV}$), the $c$-axis hoppings will be strongly
suppressed again; therefore,  NiO$_2$ and TiO$_2$ planes both develop quasi-2D
electronic states of $e_g$ and $t_{2g}$ symmetry, respectively. 
While these states are not mixed by symmetry, 
the exchange of AF spin fluctuations will lead to a sizable interplane 
coupling, having interesting implications for magnetism and possible SC in
La$_2$NiTiO$_6$.
     
The $c$-axis compression of the NiO$_6$ octahedra favoring $x^2-y^2$ orbital
can be imposed by epitaxial strain from the substrate with a lattice parameter
larger than that of LaNiO$_3$, {\it e.g.}, on LaGaO$_3$ or SrTiO$_3$ ($\simeq
3.90\text{\AA}$).  The orbital selection by tuning epitaxial strain has been
demonstrated in Ref.~\cite{Izu01}; it is based on strong Jahn-Teller response
of the $e_g$ orbital on volume conserving distortions of the oxygen octahedron
\cite{Kim00}.  In superlattices of alternating NiO$_2$ and GaO$_2$ planes
(Fig.~\ref{fig:structure}c), strain effect should be more efficient as the
lattice constants of LaGaO$_3$ and SrTiO$_3$ nearly match. Hence,
lattice relaxation is reduced and thicker superlattices La$_2$NiGaO$_6$ can be
grown. Further, using the LaGaO$_3$ substrate (instead of SrTiO$_3$) may help
to disentangle the intrinsic physics within the {\it bulk} of La$_2$NiMO$_6$
superlattice from the charge-transfer effects at the {\it interface} between
La$_2$NiMO$_6$ and the SrTiO$_3$ substrate,  
induced by different valences of Sr and La \cite{Oht02,Oka04}. 
 
Now, we turn to the theoretical examination of our proposal.  The formal
valence state in undoped case is Ni$^{3+}$ in a low-spin configuration
\mbox{$t_{2g}^6e_g^1$}, and its 4-fold degeneracy is specified by spin $S=1/2$
and orbital pseudospin $\tau=\frac12$. Because of strong $pd$-covalency, wave
functions are composed of the Ni $e_g$ states and a proper combination of the
oxygen $p$ holes of the same $e_g$ symmetry. In other words, the Ni$^{3+}$ is
shorthand notation for the \mbox{(NiO$_{6/2}$)$^{3-}$} complex with the same
-- spin doublet $\otimes$ orbital doublet -- quantum numbers.
Nearest-neighbor (NN) hopping matrix within the NiO$_2$ plane, as dictated by
symmetry, is: 
\begin{equation}\label{hopping}
t_{\alpha\beta}=
\frac{t_0}4 
\begin{pmatrix}
3 & \mp \sqrt{3} \\
\mp \sqrt{3} & 1 
\end{pmatrix} \;,\quad
\alpha,\beta\in \{x,z\}
 \;.
\end{equation}
Here, $\{x\equiv x^2-y^2,z\equiv 3z^2-r^2\}$ basis is used, $\mp$ sign is
valid for the $a$ and $b$ bond respectively. We assume small $c$-axis hopping
$\eta t_0\ll t_0$ through the block \mbox{MO$_2$} layers.

The tensile strain effect is modeled by orbital 
splitting $\Delta=\epsilon_z-\epsilon_x$. 
The (volume conserving) strain from SrTiO$_3$ substrate may provide 
$\sim 5\%$ difference between the long/short axes of NiO$_6$ 
octahedron, nearly half of that in LaMnO$_3$ where this leads 
$0.5-1.0\:\mathrm{eV}$ orbital splitting. Thus, 
$\Delta\sim0.2-0.4\:\mathrm{eV}$ might be realistic. 

The 3D nickelate \mbox{LaNiO$_3$} is a correlated two-band metal.  
Nickelates of a smaller-radii rare-earth ions, where the bandwidth 
is reduced due to a stronger GdFeO$_3$-type distortion, undergo the 
insulating state at low temperature \cite{Tor92,Alo99,Zho04,Zho05} 
and show a peculiar AF order \cite{Sca06}. 
Whether the 2D NiO$_2$ planes of La$_2$NiMO$_6$ are insulating or not 
has to be decided by experiment. Physically, the reduction of 
$c$-axis hopping and crystal-field splitting both should support the  
insulating state via the orbital disproportionation phenomenon 
in correlated systems \cite{Man02}. In other words, correlations are
effectively enhanced when the orbital degeneracy is lifted \cite{Gun96}. 
Based on these arguments, we consider below the insulating ground state. 

We derived the superexchange Hamiltonian, including both intersite ($dd$)
\mbox{Ni$^{3+}$--Ni$^{3+}$} $\rightarrow$\mbox{Ni$^{4+}$--Ni$^{2+}$} and
charge-transfer (CT) \cite{Mos04} \mbox{Ni$^{3+}-$O$^{2-}-$Ni$^{3+}$}
$\rightarrow$ \mbox{Ni$^{2+}-$O$-$Ni$^{2+}$} processes, along the lines of
Ref.~\onlinecite{Ole05}. The result for a bond $ij\parallel \gamma$ 
in the \mbox{NiO$_2$} plane can be written as 
\begin{equation}\label{Hamiltonian}
H^{(\gamma)}_{ij}=J_0(K^{dd}_{\sigma,\pm,\pm}+K^\mathrm{CT}_{\sigma,\pm,\pm})
 \Bigl(\tfrac12\pm\hat{\tau}^{(\gamma)}_i\Bigr)
 \Bigl(\tfrac12\pm\hat{\tau}^{(\gamma)}_j\Bigr) \hat{P}_{\sigma,ij}
\end{equation}
with an implied sum over $\sigma=0, 1$ and all combinations of $\pm,\pm$.
Here, $J_0=4t_0^2/U$, and $U$ stands for the Coulomb repulsion on the same
orbital. In the $c$-axis, CT part of \eqref{Hamiltonian} drops out and the
rest is multiplied by small $\eta^2$.  We have used projectors to a singlet
and triplet state of two Ni$^{3+}$ $S=1/2$ spins: 
$P_{0,ij}=\frac14-\boldsymbol{S}_i\boldsymbol{S}_j$ and
$P_{1,ij}=\frac34+\boldsymbol{S}_i\boldsymbol{S}_j$ respectively, 
and orbital projectors 
$\frac12+\tau^{(\gamma)}_i$ and $\frac12-\tau^{(\gamma)}_i$
selecting the planar orbital in the plane perpendicular to the $\gamma$-axis
and the directional orbital along this axis, {\it e.g.}, $x^2-y^2$ and
$3z^2-r^2$ orbitals, respectively, when $\gamma=c$.  While the form
\eqref{Hamiltonian} is determined by symmetry, the coefficients
$K^{dd,CT}$ are sensitive to the multiplet structure of excited states, as
usual.   Namely, they depend on two parameters $J_H/U$ and
$2U/(2\Delta_{pd}+U_p)$ characterizing the strength of the Hund coupling and
the strength of CT processes \cite{Mos04,Ole05},  respectively (explicit
expressions will be given elsewhere \cite{Cha07}).  In the following, we study
the mean-field phase diagram depending on these two parameters. 

\begin{figure}[tp]
\includegraphics[width=8.5cm]{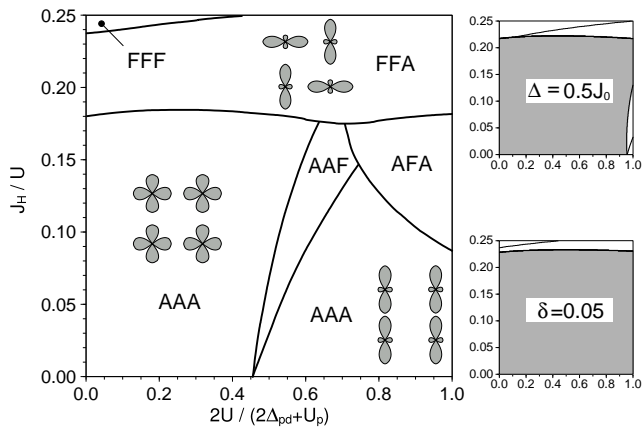}
\caption{
Phase diagram of the spin-orbital Hamiltonian \eqref{Hamiltonian}. The
$c$-axis hopping is $0.3t_0$.  The labels AAA {\it etc} indicate the
magnetic order, with F (A) denoting parallel (antiparallel) orientations of
the NN spins in the $a$, $b$ and $c$ directions, respectively. Representantive
orbital states are sketched. In the pure
planar-orbital phase (left-bottom), a weak A-coupling between the layers is
due to the higher order hoppings.  Upper right panel: the phase diagram for
$\Delta=\epsilon_z-\epsilon_x=0.5J_0$, the gray area indicates the
planar-orbital/spin-AAA phase.  Lower right panel: the phase diagram at 5\%
hole doping ($\Delta=0$, $J=0.3t_0$) shows that charge carriers strongly
favor the planar orbital phase. 
}
\label{fig:phasediag}
\end{figure}

The orbital state 
$|3z^2-r^2\rangle\cos\frac12\theta_i+ |x^2-y^2\rangle \sin\frac12\theta_i$
is specified by the site-dependent orbital angle $\theta_i$.  We assume two
sublattices with $\theta_A$ and $\theta_B$ in each NiO$_2$ plane and different
spin arrangements and minimize superexchange $\langle H \rangle$ in this
state.  The result is the phase diagram in Fig.~\ref{fig:phasediag}.
The antiferromagnetic phase in the lower left part consists of pure $x^2-y^2$
planar orbitals (as in cuprates). At higher strength of CT processes, 
this phase changes to the mixed orbital phases AAF, AAA, and AFA
($\theta_A=\theta_B$ evolve gradually from $\pi$ to $\approx\pi/2$)
with strongly anisotropic spin couplings. Upper in-plane 
ferromagnetic phases have staggered orbital order 
$\theta_A=2\pi-\theta_B\approx \pi/2$. 

Effect of the strain-induced field $\Delta$ on phase diagram is dramatic.
Even a very modest splitting $\Delta$ makes the planar orbital phase a
dominant one (Fig.~\ref{fig:phasediag}). 

% Fig.2 -------------------------------------------------------------------
%--------------------------------------------------------------------------

Next, we address the effect of mobile holes (introduced either by an
oxygen-excess or by Sr doping). Evaluating \eqref{hopping} in our
orbital state specified by $\{\theta_i\}$, we arrive at the effective 
NN hopping $\langle t_{\alpha\beta} \rangle_{ij}=
t_0\left(2\cos\theta_{-} - \cos\theta_{+} \mp
\sqrt{3}\sin\theta_{+} \right)/4$, 
where $\theta_{\pm}=(\theta_i \pm \theta_j)/2$.  
({\it E.g.}, in the planar orbital phase with $\theta=\pi$, one
finds $\langle t\rangle=3t_0/4$ as in cuprates.)  For a simple estimate,
we add kinetic energy 
$-2|\langle t\rangle_a|-2|\langle t\rangle_b|$ 
per hole to the
superexchange energy. The modified phase diagram in Fig.~\ref{fig:phasediag}
shows that the planar-orbital phase gains the largest kinetic energy
\cite{Mac99} and quickly spreads with $\delta$. 

The spin-exchange constant in the planar orbital phase takes the familiar form
(at $J_H/U=0$) \cite{Mos04}: 
$J=(9t_0^2/4)\left[1/U+2/(2\Delta_{pd}+U_p)\right]$.
With $t=3t_0/4$ (hopping between the planar orbitals), the prefactor reads
$4t^2$ as in cuprates. The value of $J$ is not much sensitive to the strength
of the Hund coupling, it is about 10\% smaller at finite $J_H/U=0.15$.

At larger doping, the kinetic energy dominates and the spin order is lost. 
We show in the following that the orbital degeneracy is lifted again and 
the planar orbital is selected by correlated motion of holes. To address this
regime, we adopt the multiorbital Gutzwiller approximation as described,
{\it e.g.}, in Ref.~\onlinecite{Lee05}. The hopping matrix elements are
renormalized $t_{\alpha\beta}\rightarrow g_{\alpha\beta} t_{\alpha\beta}$
according to the orbital occupations via
$g_{\alpha\beta}=2\delta/\sqrt{(2-n_\alpha)(2-n_\beta)}$ 
(where $\alpha,\beta\in \{x,z\}$) 
and orbital splitting is adjusted $\Delta\rightarrow\tilde{\Delta}$
to minimize energy. The renormalized dispersion
$\varepsilon_{\boldsymbol{k}}$ is then 
\begin{equation}
\frac{2\varepsilon_{\boldsymbol{k}} }{t_0} = -g_{+}\gamma_{+} \pm\sqrt{ 
\left(g_{-}\gamma_{+} +\tilde{\Delta}/t_0\right)^2+12g_{xz}^2\gamma_{-}^2 }
\label{dispersion}
\end{equation}
apart from a $\boldsymbol{k}$-independent constant. Here   
$g_\pm=3g_{xx}\pm g_{zz}$, and
$\gamma_\pm=\frac12(\cos k_a\pm\cos k_b)$.

% Fig.3 -------------------------------------------------------------------
\begin{figure}[tbp]
\includegraphics[width=7.5cm]{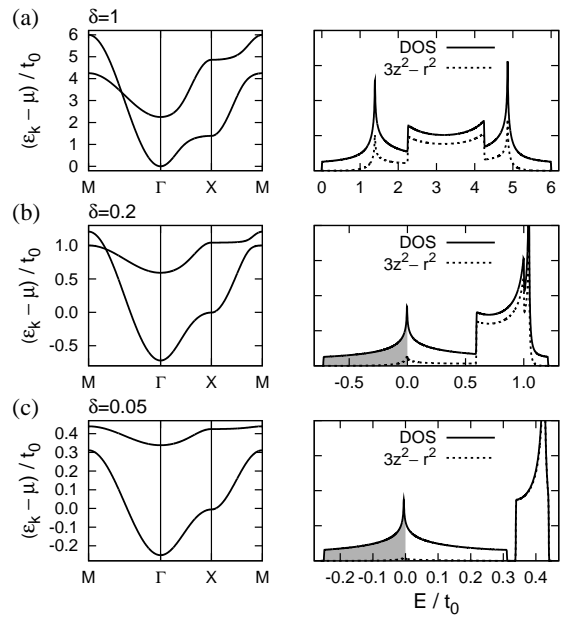}
\caption{
The band structure and density of states for the $e_g$ orbital splitting
$\epsilon_z-\epsilon_x=0.25t_0$ 
at doping levels (a) 100\% (bare bands), (b) 20\%, and (c) 5\%.  The
dotted curves show the density of states projected onto the \mbox{$3z^2-r^2$}
orbital. The gray areas indicate the occupied states at given $\delta$.
}
\label{fig:bands}
\end{figure}
%--------------------------------------------------------------------------

\begin{figure}[tbp]
\includegraphics[width=7.7cm]{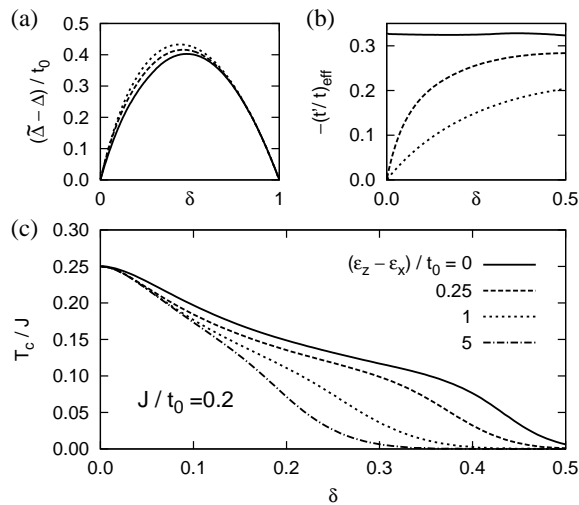}
\caption{
(a) The correlation-induced correction to the orbital splitting, and (b) the
effective $t'/t$ as function of doping~$\delta$ for different
$\epsilon_z-\epsilon_x$.  (c) The mean-field $T_c$ \cite{note1} at various
levels of $e_g$ orbital splitting.  The dashed-dotted line
($\epsilon_z-\epsilon_x=5t_0$) corresponds to the ``cuprate'' situation with
inactive  \mbox{$3z^2-r^2$} state. 
}
\label{fig:RVBtemp}
\end{figure}

Fig.~\ref{fig:bands} shows bands $(\varepsilon_{\boldsymbol{k}}-\mu)/t_0$ 
and density of states for different dopings at bare orbital splitting 
$\Delta=0.25t_0$. Correlations separate the bands and suppress the orbital 
mixing, such that the lower band is dominated by \mbox{$x^2-y^2$} orbital 
and contains only a negligible $3z^2-r^2$ fraction. Correlation-induced 
increase in orbital splitting (see Fig.~\ref{fig:RVBtemp}a) 
and a reduced mobility of the $3z^2-r^2$ orbital cooperate 
to lift the orbital degeneracy. 
The remaining effect of the initially degenerate $3z^2-r^2$ orbital is 
a deformation of the lower band of $x^2-y^2$ symmetry, 
so that the van Hove singularity stays close to the Fermi level. 

The Fermi surface shape can exactly be reproduced by an effective  $t-t'$
model with the dispersion relation $-2t(\cos k_a+\cos k_b)-4t'\cos
k_a\cos k_b$. The finite $t'/t$ values (Fig.~\ref{fig:RVBtemp}b) are
generated solely by the presence of the second band, as we started with
NN-only hopping. 

Having a single band operating near the Fermi-level, we may consider $t-t'-J$
model as in cuprates. To see effects of strain-induced splitting $\Delta$ (via
$\varepsilon_{\boldsymbol{k}}$) on possible SC, we estimated the $T_c$ values
within the mean-field treatment of $t-J$ model 
(see, {\it e.g.}, Ref.~\onlinecite{Zha88}), {\it i.e.} from   
\begin{equation}\label{Tceq}
1=2J\sum_{\boldsymbol{k}} 
\frac{\gamma_-^2}{|\varepsilon_{\boldsymbol{k}}-\mu|}
\tanh\frac{|\varepsilon_{\boldsymbol{k}}-\mu|}{2T_c} \;.
\end{equation}
We adopted here $J=0.3t \approx 0.2t_0$ (a reasonable value for the planar
orbitals), and $\varepsilon_{\boldsymbol{k}}$ is the lower branch of
Eq.~\eqref{dispersion}. 
The resulting $T_c$ curves are shown in Fig.~\ref{fig:RVBtemp}. 
As already observed in Fig.~\ref{fig:bands}, the
presence of virtual $3z^2-r^2$ orbital deforms the band dispersion 
and enhances the density of states near the Fermi level. 
The corresponding $T_c$ enhancement is clearly visible in 
Fig.~\ref{fig:RVBtemp}.

% Fig.4 -------------------------------------------------------------------
%--------------------------------------------------------------------------

As a final remark, superlattices with two subsequent NiO$_2$ planes, {\it
e.g.} ../NiNi/Ga/NiNi/.. imitating double-layer cuprates should be
interesting. As the planar orbital ordering is quite robust, a breakdown of
quasi 2D single-band picture is expected to occur at some larger critical
number of the subsequent NiO$_2$ planes \cite{Cav91}. 
  
To conclude, we pointed out that the orbitally-nondegenerate spin one-half
electronic structure -- as in cuprates -- is expected in the Ni-based
superlattices. This suggests that artificially tailored superlattices may open
new perspectives for the high-$T_c$ superconductivity.  We hope that the
theoretical expectations for our particular proposal -- ``the double 
perovskite'' La$_2$NiMO$_6$ -- are encouraging enough to motivate 
experimental efforts.   

We would like to thank B. Keimer, J. Chakhalian, O.K. Andersen, 
and P. Horsch for stimulating discussions.


\begin{thebibliography}{99}
\bibitem{note1} To be understood as a ``pseudogap temperature'' at small
doping $\delta$. To estimate the true SC temperature one has to multiply 
the $T_c$ curves by $g \approx 2\delta$ \cite{Zha88}.
\end{thebibliography}
\end{document}